\RequirePackage[pagewise,mathlines]{lineno} 
\documentclass[aps,prc,fleqn,floatfix,twocolumn,numerical,linenumbers,superscriptaddress,showpacs,twoside]{revtex4-1}
\usepackage[dvipdfm]{graphicx}
\usepackage{amsmath,amssymb,amsfonts}
\usepackage{color}

\def\v2{\mbox{$v_2$}}

\newcommand{\mean}[1]{\left\langle #1 \right\rangle}

\begin{document}


%
\title{ Initial eccentricity fluctuations and their relation to higher-order flow harmonics 
}
%
%
%
\author{ Roy~A.~Lacey}
\email[E-mail: ]{Roy.Lacey@Stonybrook.edu}
\affiliation{Department of Chemistry, 
Stony Brook University, \\
Stony Brook, NY, 11794-3400, USA}
\affiliation{Physics Department, Bookhaven National Laboratory, \\
Upton, New York 11973-5000, USA}
\author{Rui Wei} 
\affiliation{Department of Chemistry, 
Stony Brook University, \\
Stony Brook, NY, 11794-3400, USA}
\author{ J.~Jia}$^2$
\affiliation{Department of Chemistry, 
Stony Brook University, \\
Stony Brook, NY, 11794-3400, USA}
\affiliation{Physics Department, Bookhaven National Laboratory, \\
Upton, New York 11973-5000, USA}
\author{ N.~N.~Ajitanand} 
\affiliation{Department of Chemistry, 
Stony Brook University, \\
Stony Brook, NY, 11794-3400, USA}
\author{ J.~M.~Alexander}
\affiliation{Department of Chemistry, 
Stony Brook University, \\
Stony Brook, NY, 11794-3400, USA}
\author{A.~Taranenko}
\affiliation{Department of Chemistry, 
Stony Brook University, \\
Stony Brook, NY, 11794-3400, USA} 



\date{\today}


\begin{abstract}

	Monte Carlo simulations are used to compute the centrality dependence of 
the participant eccentricities ($\varepsilon_{n}$) in Au+Au collisions, for 
the two primary models currently employed for eccentricity estimates --
the Glauber and the factorized Kharzeev-Levin-Nardi (fKLN) models.
They suggest specific testable predictions for the magnitude and centrality 
dependence of the flow coefficients $v_n$, respectively measured relative 
to the event planes $\Psi_n$. They also indicate that the ratios of several 
of these coefficients may provide an additional constraint for distinguishing 
between the models. Such a constraint could be important for a more precise 
determination of the specific viscosity of the matter produced in heavy ion collisions.
 
\end{abstract}


\maketitle


	Collective flow continues to play a central role in ongoing efforts to 
characterize the transport properties of the strongly interacting matter produced 
in heavy ion collisions at the Relativistic Heavy Ion Collider 
(RHIC) \cite{Gyulassy:2004zy,Molnar:2004yh,Lacey:2006bc,Adare:2006nq,Romatschke:2007mq,
Xu:2007jv,Drescher:2007cd,Shuryak:2008eq,Luzum:2008cw,Song:2008hj,Chaudhuri:2009hj,
Xu:2007jv,Dusling:2007gi,Greco:2008fs,Bozek:2009mz,Denicol:2010tr,Lacey:2010fe}. 
An experimental manifestation of this flow is the anisotropic emission of particles 
in the plane transverse to the beam direction \cite{Lacey:2001va,Snellings:2001nf}. 
This anisotropy can be characterized by the even order Fourier coefficients;
\begin{equation}
v_{\rm n} = \mean{e^{in(\phi_p - \Psi_{RP})}}, {\text{  }} n=2,4,.., 
\label{eq:1}
\end{equation}  
where $\phi_{p}$ is the azimuthal angle of an emitted particle, 
$\Psi_{RP}$ is the azimuth of the reaction plane and the brackets denote 
averaging over particles and events \cite{Ollitrault:1992bk}. Characterization 
has also been made via the pair-wise distribution in the azimuthal angle difference 
($\Delta\phi =\phi_1 - \phi_2$) between particles \cite{Lacey:2001va,Adcox:2002ms,Mocsy:2010um};
\begin{equation}
\frac{dN^{\text{pairs}}}{d\Delta\phi} \propto \left( 1 + \sum_{n=1}2v_n^2\cos(n\Delta\phi) \right).
\label{eq:2}
\end{equation}

	Anisotropic flow is understood to result from an asymmetric hydrodynamic-like expansion
of the medium produced by the two colliding nuclei. That is, the spacial asymmetry of the 
produced medium drives uneven pressure gradients in- and out of the reaction plane and hence, 
a momentum anisotropy of the particles emitted about this plane. This mechanistic 
picture is well supported by the observation that the measured anisotropy for hadron $p_T \alt 2$ GeV/c, 
can be described by relativistic hydrodynamics \cite{Heinz:2001xi,Teaney:2003kp,Huovinen:2001cy, 
Hirano:2002ds,Andrade:2005tx,Nonaka:1999et,Romatschke:2007mq,Song:2008hj,Niemi:2008ta,
Dusling:2007gi,Bozek:2009mz,Peschanski:2009tg,Denicol:2010tr,Holopainen:2010gz,Schenke:2010rr}. 

	The differential Fourier coefficients $v_{2}(N_{\text part})$ 
and $v_{2}({p_T})$ have been extensively studied in Au+Au collisions 
at RHIC \cite{Adcox:2002ms,Adams:2003zg,Adler:2003kt,Alver:2006wh,
Afanasiev:2007tv,Star:2008ed,Afanasiev:2009wq,Adare:2010ux}. 
One reason for this has been the realization that these elliptic flow 
coefficients are sensitive to various transport properties of the 
expanding hot medium \cite{Heinz:2002rs,Teaney:2003kp,Lacey:2006pn,
Romatschke:2007mq,Song:2007ux,Drescher:2007cd,Xu:2007jv,Greco:2008fs,
Luzum:2008cw,Chaudhuri:2009hj}.
Indeed, considerable effort has been, and is being devoted to the quantitative extraction 
of the specific shear viscosity $\eta/s$ ({\em i.e.}\ the ratio 
of shear viscosity $\eta$ to entropy density $s$) via comparisons to viscous relativistic 
hydrodynamic simulations \cite{Luzum:2008cw,Song:2008hj,Dusling:2007gi,
Chaudhuri:2009hj,Bozek:2009mz,Denicol:2010tr,Holopainen:2010gz}, transport model 
calculations \cite{Molnar:2001ux,Xu:2007jv,Greco:2008fs} and hybrid approaches which 
involve the parametrization of scaling violations to ideal hydrodynamic behavior \cite{Drescher:2007cd,Lacey:2006pn,Lacey:2009xx,Masui:2009pw,Lacey:2010fe}. 
The initial eccentricity of the collision zone and its associated fluctuations, has proven to be an essential ingredient for these extractions.

	Experimental measurements of the eccentricity have not been possible to date. Consequently, much reliance has been placed on the theoretical estimates obtained 
from the overlap geometry of the collision zone, specified by the impact parameter $b$ or the number of participants $N_{\text{part}}$
\cite{Miller:2003kd,Alver:2006wh,Hama:2007dq,Broniowski:2007ft,
Hirano:2009ah,Lacey:2009xx,Gombeaud:2009ye,Lacey:2010yg,Schenke:2010rr,Staig:2010pn,Qin:2010pf}. 
For these estimates, the geometric fluctuations associated 
with the positions of the nucleons in the collision zone, serve as the 
underlying cause of the initial eccentricity fluctuations. That is, 
the fluctuations of the positions of the nucleons lead to fluctuations of 
the so-called participant plane (from one event to another)
which result in larger values for the eccentricities ($\varepsilon$)
referenced to this plane.

	The magnitude of these fluctuations are of course model dependent, and this  
leads to different predictions for the magnitude of the eccentricity.  
More specifically, the $\varepsilon_2$ values obtained from the 
Glauber \cite{Alver:2006wh,Miller:2007ri} 
and the factorized Kharzeev-Levin-Nardi (fKLN) \cite{Lappi:2006xc,Drescher:2007ax} models, 
(the two primary models currently employed for eccentricity estimates) 
give results which differ by as much 
as {$\sim 25$\%} \cite{Hirano:2005xf,Drescher:2006pi} -- a difference which leads 
to an approximate factor of two uncertainty in the extracted  
$\eta/s$ value \cite{Luzum:2008cw,Lacey:2010fe}. Thus, a more precise extraction 
of $\eta/s$ requires a clever experimental technique which can measure the 
eccentricity and/or the development of experimental constraints which can 
facilitate the requisite distinction between the models used to calculate 
eccentricity.

	Recently, significant attention has been given to the study of the influence of 
initial geometry fluctuations on higher order eccentricities $\varepsilon_{n, n\geq 3}$
\cite{Broniowski:2007ft,Lacey:2010yg,Alver:2010gr,Alver:2010dn,Holopainen:2010gz,Petersen:2010cw,
Staig:2010pn,Qin:2010pf,Schenke:2010rr}, 
with an eye toward a better understanding of how such fluctuations manifest into 
the harmonic flow correlations characterized by $v_n$ (for odd and even $n$), and whether they 
can yield constraints that could serve to pin down the ``correct'' model for eccentricity
determination. For the latter, the magnitude of $\varepsilon_n$ and its detailed 
centrality dependence is critical. Therefore, it is essential to resolve the substantial differences in the $\varepsilon_{n}$ values reported and used by different authors
\cite{Broniowski:2007ft,Lacey:2010yg,Alver:2010gr,Alver:2010dn,Holopainen:2010gz,Petersen:2010cw,Staig:2010pn,Qin:2010pf,Schenke:2010rr}.

	Here, we argue that the magnitudes and trends for the eccentricities 
$\varepsilon_{n}$ imply specific testable predictions for the magnitude 
and centrality dependence of the flow coefficients 
$v_n$, measured relative to their respective event planes $\Psi_n$. 
We also show that the values for $\varepsilon_{n}$ obtained for the Glauber 
\cite{Alver:2006wh,Miller:2007ri} and fKLN  \cite{Lappi:2006xc,Drescher:2007ax} 
models, indicate sizable model dependent differences which could 
manifest into experimentally detectable differences in the centrality 
dependence of the ratios ${v_{3}}/{(v_{2})^{3/2}}$, ${v_{4}}/{(v_{2})^2}$ and ${v_2}/{v_{n, n\geq 3}}$. Such a constraint 
could be important for a more precise determination of the specific 
viscosity of the hot and dense matter produced in heavy ion collisions.

\section{Eccentricity Simulations}
	
	Monte Carlo (MC) simulations were used to calculate event averaged 
eccentricities (denoted here as $\varepsilon_{n}$) in Au+Au collisions,
within the framework of the Glauber (MC-Glauber) and fKLN (MC-KLN) models.
For each event, the spatial distribution of nucleons in the colliding nuclei were generated according to the Woods-Saxon function:
\begin{equation}
\rho(\mathbf{r})=\frac{\rho_{0}}{1+e^{(\text{r}-R_{0})/d}},
\label{Eq2}
\end{equation}
where $R_{0} = 6.38\,\text{fm}$ is the radius of the Au nucleus and $d=0.53\,\text{fm}$ is the 
diffuseness parameter. 

 For each collision, the values for $N_{\rm part}$ and the number of binary collisions $N_{\text{coll}}$ were 
determined within the Glauber ansatz \cite{Miller:2007ri}. 
The associated $\varepsilon_{n}$ values were then evaluated from the 
two-dimensional profile of the density of sources in the transverse  
plane $\rho_s(\mathbf{r_{\perp}})$, using modified versions 
of MC-Glauber \cite{Miller:2007ri} and MC-KLN \cite{Drescher:2007ax} respectively.

	For each event, we compute an event shape vector $S_{n}$ and the azimuth of 
the the rotation angle $\Psi_n$ for $n$-th harmonic of the shape 
profile \cite{Broniowski:2007ft,Lacey:2010yg}; 
\begin{eqnarray}  
S_{nx} & \equiv & S_n \cos{(n\Psi_n)} = 
\int d\mathbf{r_{\perp}} \rho_s(\mathbf{r_{\perp}}) \omega(\mathbf{r_{\perp}}) \cos(n\phi), \label{eq:S_x} \\
S_{ny} & \equiv & S_n \sin{(n\Psi_n)} = 
\int d\mathbf{r_{\perp}} \rho_s(\mathbf{r_{\perp}}) \omega(\mathbf{r_{\perp}}) \sin(n\phi), \label{eq:S_y} \\
    \Psi_n & = & \frac{1}{n} \tan^{-1}\left(\frac{S_{ny}}{S_{nx}}\right), \label{eq:S_n-plane}
\end{eqnarray}
where $\phi$ is the azimuthal angle of each source and 
the weight $\omega(\mathbf{r_{\perp}}) = \mathbf{r_{\perp}}^2$ and  $\omega(\mathbf{r_{\perp}}) = \mathbf{r_{\perp}}^n$ are used 
in respective calculations.
Here, it is important to note that the substantial differences reported for $\varepsilon_{n}$ in Refs. 
\cite{Broniowski:2007ft,Lacey:2010yg,Alver:2010gr,Alver:2010dn,
Holopainen:2010gz,Petersen:2010cw,Staig:2010pn,Qin:2010pf,Schenke:2010rr} is  
largely due to the value of $\omega(\mathbf{r_{\perp}})$ employed.

The eccentricities were calculated as:
\begin{eqnarray}
\varepsilon_n = \left\langle \cos n(\phi - \Psi_n) \right\rangle \,\,\,
\label{enPsin}
\end{eqnarray}
and 
\begin{eqnarray}
\varepsilon^*_n = \left\langle \cos n(\phi - \Psi_m) \right\rangle, \, \, n \ne m.
\label{e2e4}
\end{eqnarray}
where the brackets denote averaging over sources and events belonging 
to a particular centrality or impact parameter range; the starred notation is 
used here to distinguish the $n$-th order moments obtained relative 
to an event plane of a different order $\Psi_m$. 

For the MC-Glauber calculations, an additional entropy density weight was applied 
reflecting the combination of spatial coordinates of participating nucleons and binary 
collisions  \cite{Hirano:2005xf,Hirano:2009ah} ;
\begin{eqnarray}
\rho_s(\mathbf{r_{\perp}}) \propto \left[ \frac{(1-\alpha)}{2}\frac{dN_{\text{part}}}{d^2\mathbf{r_{\perp}}} + 
                     \alpha \frac{dN_{\text{coll}}}{d^2\mathbf{r_{\perp}}} \right], 
\label{Eq5}
\end{eqnarray}
where $\alpha = 0.14$ was constrained by multiplicity measurements as a 
function of $N_{\text{part}}$ for Au+Au collisions \cite{Back:2004dy}.
These procedures take account of the eccentricity fluctuations which stem 
from the event-by-event misalignment between the short axis of the ``almond-shaped'' 
collision zone and the impact parameter. 	
Note that $\varepsilon_{n}$ (cf. Eq. \ref{enPsin}) corresponds 
to $v_{n}$ measurements relative to the so-called participant 
planes \cite{Alver:2006wh,Miller:2007ri}. That is, each harmonic $\varepsilon_{n}$ 
is evaluated relative to the principal axis determined by maximizing the $n$-th moment. 
This is analogous to the measurement of $v_n$ with respect to the $n$-th 
order event-plane in actual experiments \cite{rxn_plane}. It however, contrasts 
recent  experimental measurements in which a higher order coefficient ($v_4$) 
has been measured with respect to a lower order event 
plane ($\Psi_2$) \cite{Adams:2004bi,Adare:2010ux}. Note as well that we have 
established that the angles $\Psi_n$ for the odd and even harmonics are 
essentially uncorrelated for the $N_{\text{part}}$ range of interest to this study.
%
\begin{figure}[tb]
\includegraphics[width=1.0\linewidth]{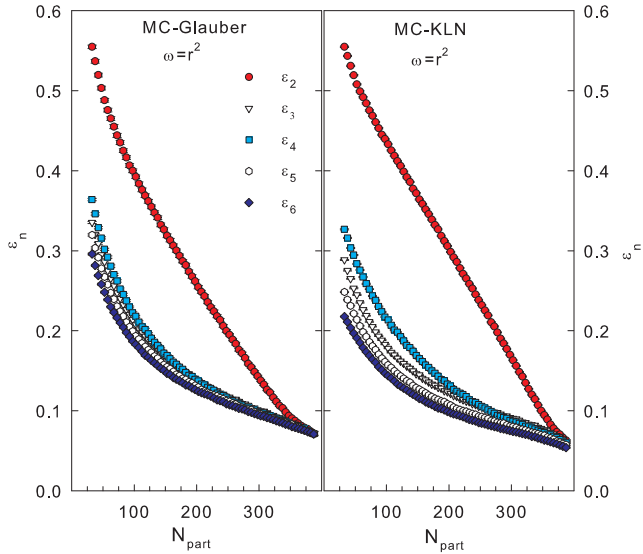}
%
\caption{ Calculated values of $\varepsilon_{n, n\leq 6}$ vs. $N_{\rm part}$
for $\omega(\mathbf{r_{\perp}}) = \mathbf{r_{\perp}}^2$
for MC-Glauber (a) and MC-KLN (b) for Au+Au collisions. The open and
filled symbols indicate the results for odd and even harmonics respectively.
}
\label{Fig1}
\end{figure}
\subsection{Results for $\omega(\mathbf{r_{\perp}}) = \mathbf{r_{\perp}}^2$ 
and $\omega(\mathbf{r_{\perp}}) = \mathbf{r_{\perp}}^n$}
      
	Figure \ref{Fig1} shows a comparison of $\varepsilon_{n, n\leq 6}$ vs. 
$N_{\text{part}}$ for $\omega(\mathbf{r_{\perp}}) = \mathbf{r_{\perp}}^2$, for 
 MC-Glauber (a) and MC-KLN (b) for Au+Au collisions. The filled and open symbols 
 indicate the results for the even and odd harmonics respectively.
For this weighting scheme, $\varepsilon_{n}$ is essentially
the same for $n \geq 3$, and have magnitudes which are 
significantly less than that for $\varepsilon_{2}$, except in very 
central collisions where the effects of fluctuation dominate the magnitude of
$\varepsilon_{n, n\geq 2}$. Note the approximate $1/\sqrt(N_{\text{part}})$
dependence for $\varepsilon_{n, n\geq 3}$. The smaller magnitudes for 
$\varepsilon_{n, n\geq 3}$ (with larger spread) apparent in Fig. \ref{Fig1}(b), 
can be attributed to the sharper transverse density distributions for MC-KLN.

Figure \ref{Fig2} shows a similar comparison of 
$\varepsilon_{n, n\leq 6}$ vs. $N_{\text{part}}$ for calculations performed
with the weight $\omega(\mathbf{r_{\perp}}) = \mathbf{r_{\perp}}^n$. 
This weighting results in an increase in the sensitivity to the outer regions of 
the transverse density distributions. Consequently, the overall magnitudes for 
$\varepsilon_{n, n\geq 3}$ are larger than those shown in Fig. \ref{Fig1}. 
This weighting also lead to a striking difference in the relative magnitudes 
of $\varepsilon_{n, n\geq 2}$ for MC-Glauber (a), MC-KLN (b) and the results for 
$\omega(\mathbf{r_{\perp}}) = \mathbf{r_{\perp}}^2$
shown in Fig. \ref{Fig1}.

\section{Eccentricity ratios}
%
\begin{figure}[t]
\includegraphics[width=1.0\linewidth]{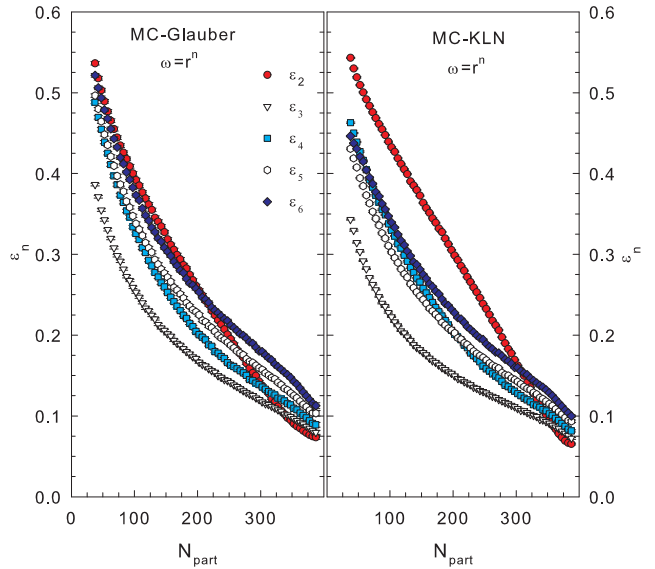}
\caption{Same as Fig. \ref{Fig1} for $\omega(\mathbf{r_{\perp}}) = \mathbf{r_{\perp}}^n$. 
}
\label{Fig2}
\end{figure}

The magnitudes and trends of the calculated eccentricities shown 
in Figs. \ref{Fig1} and \ref{Fig2} are expected to 
influence the measured values of $v_{n}$. To estimate this influence, 
we first assume that the resulting anisotropic flow is directly proportional
to the initial eccentricity, as predicted by perfect fluid hydrodynamics. Here,
our tacit assumption is that a possible influence from the effects of a 
finite viscosity ($\eta/s$) is small because current estimates indicate 
that $\eta/s$ is small \cite{Lacey:2006pn,Adare:2006nq,Drescher:2007cd,Dusling:2007gi,
Xu:2007jv,Luzum:2008cw,Song:2008hj,Greco:2008fs,Chaudhuri:2009hj,Lacey:2009xx,
Masui:2009pw,Bozek:2009mz,Denicol:2010tr,Lacey:2010fe,Holopainen:2010gz} -- of the same  
magnitude as for the conjectured KSS bound $\eta/s = 1/(4\pi)$ \cite {Kovtun:2004de}. 

%
\begin{figure}[!h]
\includegraphics[width=1.0\linewidth]{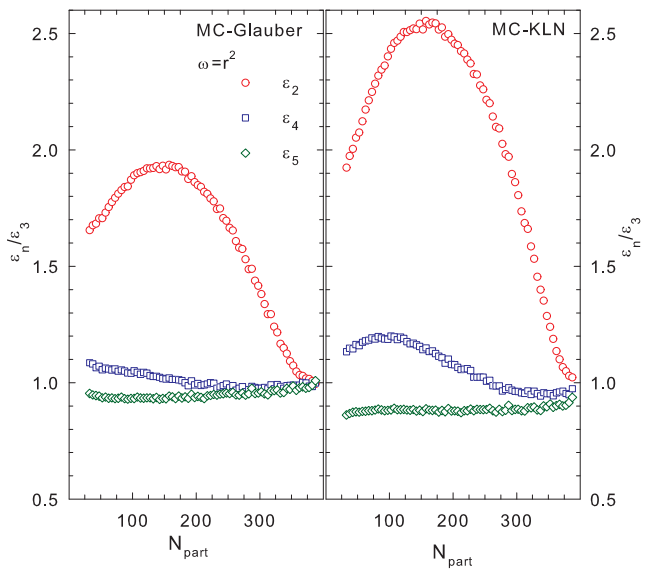}
\caption{Comparison of ${\varepsilon_{2,4,5}}/{\varepsilon_{3}}$ vs. $N_{\text{part}}$ for Au+Au collisions. Results are shown for MC-Glauber (a)
and MC-KLN (b) calculations. 
}
\label{Fig3}
\end{figure}

Figure \ref{Fig1} indicates specific testable predictions for the relative 
influence of $\varepsilon_{n,n\geq 2}$ on the magnitudes of $v_{n,n\geq 2}$. 
That is, (i) $\varepsilon_2$ should have a greater influence than 
$\varepsilon_{n,n\geq 3}$ in non-central collisions, (ii) the respective 
influence of $\varepsilon_{n,n\geq 3}$ on the values for $v_{n,n\geq 3}$ should 
be similar irrespective of centrality and (iii) the ratios $v_{4,5,6}/v_3$ 
should follow a specific centrality dependence due to the influence of 
$\varepsilon_{4,5,6}/\varepsilon_3$. Such a dependence is illustrated in Fig. \ref{Fig3} 
where we show the centrality dependence of the ratios $\varepsilon_{2,4,5}/\varepsilon_3$, 
obtained for MC-Glauber (a) and MC-KLN (b) calculations.
They suggest that, if MC-Glauber-like eccentricities, with weight  
$\omega(\mathbf{r_{\perp}}) = \mathbf{r_{\perp}}^2$, are the relevant 
eccentricities for Au+Au collisions, then the measured ratio $v_2/v_3$ 
should increase by a factor $\approx 2$, from central to mid-central 
collisions ($N_{\text{part}} \sim 350-150$). 
For $N_{\text{part}} \alt 150$, Fig. \ref{Fig2}(a) shows that 
the ratio $v_2/v_3$ could even show a modest decrease. 
The eccentricity ratios involving the higher harmonics suggest that, if they 
are valid, the measured values of $v_{4,5,6}/v_3$ should show little, if any, 
dependence on centrality, irrespective of their magnitudes.

	The ratios $\varepsilon_{2,4,5}/\varepsilon_3$ obtained for MC-KLN calculations 
are shown in Fig. \ref{Fig3} (b). While they indicate qualitative trends which are similar 
to the ones observed in Fig. \ref{Fig3} (a), their magnitudes and their detailed dependence 
on centrality are different. Therefore, if the qualitative trends discussed earlier were 
indeed found in data, then these differences suggest that precision measurements of the 
centrality dependence of the relative ratios for ${v_2}/{v_3}$, ${v_4}/{v_3}$,
${v_5}/{v_3}, ...$ for several $p_T$ selections, could provide a constraint for aiding 
the distinction between fKLN-like and Glauber-like initial collision geometries. 
Specifically, smaller (larger) values of the relative ratios are to be expected for 
${v_2}/{v_3}$ and ${v_4}/{v_3}$ for Glauber-like (fKLN-like) 
initial geometries. Note the differences in the expected centrality dependencies as well.
%
\begin{figure}[t]
\includegraphics[width=1.0\linewidth]{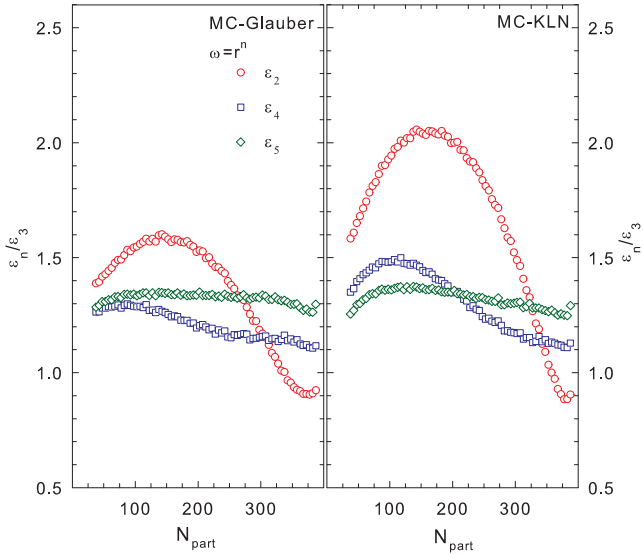}
\caption{Same as Fig. \ref{Fig3} for $\omega(\mathbf{r_{\perp}}) = \mathbf{r_{\perp}}^n$.
}
\label{Fig4}
\end{figure}

	Figure \ref{Fig4} compares the eccentricity ratios $\varepsilon_{2,4,5}/\varepsilon_3$ 
obtained for MC-Glauber (a) and MC-KLN (b) calculations with the weight 
$\omega(\mathbf{r_{\perp}}) = \mathbf{r_{\perp}}^n$. The magnitudes of these ratios and their 
centrality dependencies are distinct for MC-Glaber and MC-KLN. They are also 
quite different from the ratios shown in Fig. \ref{Fig3}. This suggests that    
precision measurements of the centrality dependence of the relative ratios 
${v_2}/{v_3}$, ${v_4}/{v_3}$, 
${v_5}/{v_3}, ...$ (for several $p_T$ selections) should not only allow 
a clear distinction between  MC-Glauber and MC-KLN initial geometries, but also 
a distinction  between the the $\omega(\mathbf{r_{\perp}}) = \mathbf{r_{\perp}}^2$ and
$\omega(\mathbf{r_{\perp}}) = \mathbf{r_{\perp}}^n$ weighting methods.


   A finite viscosity will influence the magnitudes of $v_n$. Thus, 
for a given $p_T$ selection, the measured ratios for ${v_2}/{v_3}$, 
${v_4}/{v_3}$, ${v_5}/{v_3}, ...$ will be different 
from the eccentricity ratios shown in Figs. \ref{Fig3} and \ref{Fig4}. 
Note as well that, even for ideal hydrodynamics, the predicted magnitude of 
${v_4}/{\varepsilon_4}$ is only a half of that for 
${v_2}/{\varepsilon_2}$ \cite{Alver:2010dn}. 
Nonetheless, the rather distinct centrality dependent eccentricity patterns exhibited in 
Figs. \ref{Fig3} and \ref{Fig4} suggests that measurements of the ratios 
of these flow harmonics should still allow a distinction between  MC-Glauber and MC-KLN 
initial geometries, as well as a distinction between the two weighting methods.
%
\begin{figure}[t]
\includegraphics[width=1.0\linewidth]{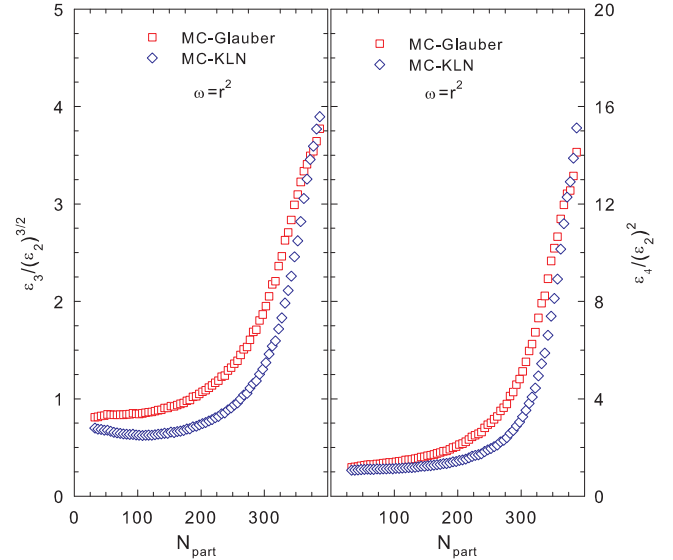} 
\caption{Comparison of $\varepsilon_3/(\varepsilon_2)^{3/2}$ vs. $N_{\text{part}}$ (a) 
and  $\varepsilon_4/(\varepsilon_2)^{2}$ vs. $N_{\text{part}}$ (b) for MC-Glauber 
and MC-KLN initial geometries (as indicated) for Au+Au collisions.
}
\label{Fig5}
\end{figure}

\begin{figure}[t]
\includegraphics[width=1.0\linewidth]{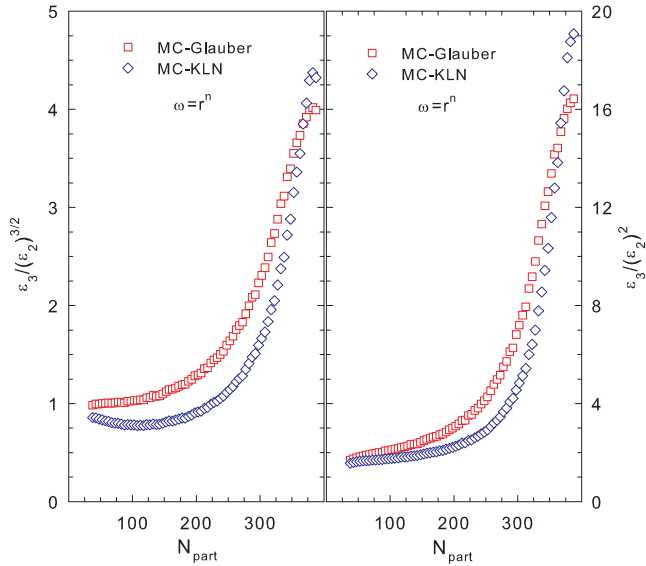} 
\caption{Same as Fig. \ref{Fig5} for $\omega(\mathbf{r_{\perp}}) = \mathbf{r_{\perp}}^n$.
}
\label{Fig6}
\end{figure}

	The ratios $v_3/(v_2)^{3/2}$ and $v_4/(v_2)^{2}$ have been recently found to scale 
with $p_T$ \cite{Lacey:WWND2011}, suggesting  a reduction in the influence of viscosity on  
them. Thus, the measured ratios $v_n/(v_2)^{n/2}$ could give a more 
direct indication of the centrality dependent influence of 
$\varepsilon_n/(\varepsilon_2)^{n/2}$ on $v_n/(v_2)^{n/2}$.
The open symbols in Figs. \ref{Fig5} and \ref{Fig6} indicate a substantial 
difference between the ratios $\varepsilon_3/(\varepsilon_2)^{3/2}$ (a) 
and $\varepsilon_4/(\varepsilon_2)^{2}$ (b) for the MC-Glauber and MC-KLN 
geometries as indicated. 
Note as well that the ratios in Fig. \ref{Fig6} are substantially larger than those 
in Fig. \ref{Fig5}. The latter difference reflects the different weighting schemes 
used, i.e. $\omega(\mathbf{r_{\perp}}) = \mathbf{r_{\perp}}^n$ and 
$\omega(\mathbf{r_{\perp}}) = \mathbf{r_{\perp}}^2$ respectively.
Interestingly, the ratios for $\varepsilon_4/(\varepsilon_2)^{2}$
imply much larger measured ratios for ${v_{4}}/{(v_{2})^2}$ than the 
value of 0.5 predicted by perfect fluid hydrodynamics (without fluctuations)
\cite{Borghini:2005kd,Csanad:2003qa}. However, they show qualitative trends which 
are similar to those for the measured ratios $v_4/(v_2)^2$, obtained for $v_4$ evaluations 
relative to the $\Psi_2$ plane\cite{Adams:2004bi,Adare:2010ux}. 
The relatively steep rise of the ratios in Figs. \ref{Fig5} and \ref{Fig6} 
(albeit steeper for MC-Glauber), can be attributed to the larger influence that 
fluctuations have on the higher harmonics. Note that these are the same fluctuations 
which give rise to the ``anomalously low'' values 
of $\varepsilon_4$ evaluated with respect to $\Psi_2$ in central collisions \cite{Lacey:2010yg}. 

Figures \ref{Fig3} - \ref{Fig6} suggests that measurements of the centrality dependence of the ratios ${v_{3}}/{(v_{2})^{3/2}}$ and ${v_{4}}/{(v_{2})^2}$, in conjunction with those for 
${v_2}/{v_3}$, ${v_4}/{v_3}$, ${v_5}/{v_3}...$ may provide 
a robust constraint for the role of initial eccentricity fluctuations, 
as well as an additional handle for making a distinction between 
Glauber-like and fKLN-like initial geometries. These measurements could also lend insight, as well as place important constraints for the degree to which a small value of $\eta/s$ and/or the effects of thermal smearing, modulate the higher order flow harmonics [compared to $v_2$] as has been 
suggested \cite{Petersen:2010cw,Qin:2010pf,Schenke:2010rr}.
%

\section{Summary}

In summary, we have presented results for the initial eccentricities 
$\varepsilon_{n, n\leq 6}$ for Au+Au collisions  with different 
weighting schemes, for the two primary models currently employed 
for eccentricity estimates at RHIC. The calculated values of $\varepsilon_{n, n\leq 6}$,
which are expected to influence the measured flow harmonics $v_n$,  
suggests that measurements of the centrality dependence 
of ${v_2}/{(v_3)}$, ${v_4}/{v_3}$, ${v_{3}}/{(v_{2})^{3/2}}$, ${v_{4}}/{(v_{2})^2}$, etc.
could provide stringent constraints for validating the predicted influence 
of eccentricity fluctuations on $v_n$, as well as an important additional 
handle for making a distinction between Glauber-like and fKLN-like initial 
geometries. Measurements of $v_n$ and their ratios are 
now required to exploit these simple tests. 

{\bf Acknowledgments}
We thank Wojciech Broniowski for profitable discussions and invaluable 
model calculation cross checks. This research is supported by the US DOE under contract DE-FG02-87ER40331.A008 and by the NSF under award number PHY-1019387.
 


%
\bibliography{ecc_fluc_x0} 

\begin{thebibliography}{67}
\bibitem [{rxn()}]{rxn_plane}%
  \BibitemOpen
  \href@noop {} {}\bibinfo {note} {Note that the event planes for the
  eccentricities are specified by the initial state coordinate asymmetry,
  whereas the experimental event planes for $v^*_n$ are specified by the final
  state momentum space anisotropy. The two planes are correlated in ideal
  hydrodynamics.}\BibitemShut {Stop}%
\end{thebibliography}
\end{document}